# A Robotic Framework for Making Eye Contact with Humans

**Mohammed Moshiul Hoque**
Department of Computer Science and Engineering
Chittagong University of Engineering and Technology
Chittagong, Bangladesh. E-mail: moshiul_240@cuet.ac.bd

**Abstract:** *Meeting eye contact is a most important prerequisite skill of a human to initiate any conversation with others. However, it is not easy task for a robot to meet eye contact with a human if they are not facing each other initially or the human is intensely engaged his/her task. If the robot would like to start communication with a particular person, it should turn its gaze to that person first. However, only such a turning action alone is not always be enough to set up eye contact. Sometimes, the robot should perform some strong actions so that it can capture the human's attention toward it. In this paper, we proposed a computational model for robots that can pro-actively captures human attention and makes eye contact with him/her. Evaluation experiment by using a robotic head reveals the effectiveness of the proposed model in different viewing situations.*

## 1. Introduction

Currently work in robotics is expanding from industrial robots to robots that are employed in the living environment. For robots to be accepted into the real world, they must be capable to behave in such a way that humans do with other humans. Although a number of significant challenges remained unsolved related to the social capabilities of robots, the robot that can pro-actively meets eye contact with human is also an important research issue in the realm of natural HRI.

Eye contact is a phenomenon that occurs when two people cross their gaze which plays an important role in initiating an interaction and in regulating face-to-face communication [1, 29]. Eye contact behaviour is the basis of and developmental precursor to more complex gaze behaviours such as joint visual attention [7], turn-taking [30], information recall [8], and so on. For any social interaction to be initiated and maintained, parties need to establish eye contact [10]. However, it is very difficult to establish such gaze behaviours for one person while the target person is not facing him/her or while target people are intensely attending his/her task.

A robot that naturally makes eye contact with human is one of its major capabilities to be implemented in social robots. Capturing attention and ensuring while capturing attention are the two important prerequisites for making an eye contact episode. After capturing the attention of the intended recipient, the robot needs to make the person notice clearly that it is looking at none other than him/her. In order to create awareness explicitly, the robot should display some actions (i.e, facial expression, nodding, and so on).

Situation where the human and the robot are not facing each other initially needs robots use a proactive approach to the intended human for making eye contact. This approach enables robots to help people who have potential needs and convey some information about an object or a particular direction that the human should focus. In summary, the major issues in our research are: (i) how can a robot use subtle cues to capture the human's attention if s/he is not facing to the robot, in other words, if the robot cannot capture his/her eyes or whole face due to the spatial arrangements of the person and the robot, and (ii) how robot ensure that the human is responding against its action and how it tell when it has captured attention? To answer these issues we proposed a framework and we design a robotic head based on this that confirmed as effective to make eye contact with the human in experimental evaluation.

## 2. Hypotheses in Making Eye Contact

Humans usually turn their head or gaze first toward the person with whom they would like to communicate [23]. If the target human does not respond, s/he tries again with the same action or with the more strong signals (e.g., waving hand, shaking head, moving body, or voice, etc.). Robots should use the same convention as humans in a natural HRI scenario. Attention capture can produce observable behavioural responses such as eye, head movements, or body orientation [26]. Therefore, if the target person felt attracted by the robot, s/he will turn toward it, which will make face-to-face orientation (i.e., gaze crossing of each other). Psychological studies show however, that this gaze crossing action alone may not be enough to make a successful eye contact event [8]. That means, the robot needs to make the person notice clearly that it is looking at none other than him/her. In to display awareness explicitly the robot should use some actions (verbal or non-verbal).

Based on the above discussion, we can hypotheses that robots should perform two tasks consecutively: (i) attention capture, and (ii) ensuring attention capture for making eye contact pro-actively. Figure 1 illustrates the conceptual process of attention attraction in terms of these tasks. To perform a successful eye contact episode, both a robot (R) and a human (H) need to show some explicit behaviours and to respond appropriately to them by communicative behaviours in each phase. That means, R and H performs a set of behaviours, R= $\{\alpha,\delta\}$ and H= $\{\lambda,\theta\}$ respectively.

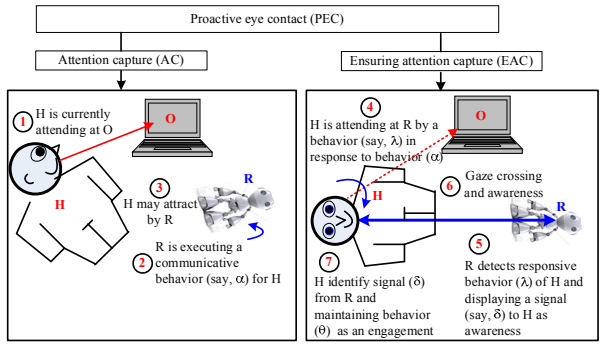

Figure 1. Prerequisites of making eye contact proactively.

In this work, we apply a set of behaviours of robot such as, $\alpha$ = *{head_turning, head_shaking, reference_terms}* in attention capture phase, and $\delta$ = *{face-detection, eye_blinks}* in ensuring attention capture phase. We are also expecting human's behaviours such as, $\lambda=\{(head \wedge gaze \wedge body)$ *turn_toward_robot}* in attention capture phase, and $\theta$ = *{keep_looking_toward_robot}* in ensuring attention capture phase.

## 3. Related Work

Several previous HRI studies have addressed greeting behavior to initiates human-robot conversation and eye contact process at a social distance. These robots are designed to utter some greeting terms to initiates the interaction with the human [11, 13, 24]. Some robots were equipped with the capability to encourage people to make eye contact by some non-verbal cues such as body orientation and gaze [18], approaching direction [5], standing position [32], following behaviours [12]. Robots may wait for a person to initiate an interaction process. These studies assumed that the target person faces the robot and intends to talk with it; however, in actual practice this assumption may not always hold. Although such a passive attitude can work in some situations, many situations require a robot to employ a more active approach [3, 25, 28].

Some robots were equipped with the capability to initiate interaction pro-actively with humans [27, 19]. Their systems fails to recognize people's gaze direction, which is the most important parameter to measure whether the people have responded (been attracted) to the robot's intentional signal or not. Several others robotic systems were developed to establish eye contact [16, 22]. These robots are supposed to make eye contact with humans by turning their eyes (cameras) toward the human faces. All of these studies focus only on the gaze crossing function of the robots as making its eye contact capability and gaze awareness functions are absent.

Several robotic systems were incorporates gaze awareness functions too. For example, Miyauchi et al. [20] design a system that can make eye contact between human and robot. This robot used a flat screen monitor as the robot's head and display 3D computer graphics (CG) images to produce smiling expression as gaze awareness function. A flat screen is unnatural as a face. Yoshikawa et al. [32] used a communication robot to produce the responsive gaze behaviors of the robot. However, the robotic head that used in this study was mechanically very complex and as such expensive to design, construct and maintain. A recent work that used a robot *Simon* to produce the awareness function [14] by blinks its ear. Although they consider the single person interaction scenario, they did not used ear blinks as a gaze awareness purpose rather use to create interaction awareness.

## 4. System Architecture

We have developed a robotic head for HRI experiments. In the following sections, we discuss the architecture of our robotic systems and its behaviours in details.

### 4.1 Hardware Configuration

Figure 2 shows an overview of our robotic head. The head consists of a spherical 3D mask, an LED projector (3M pocket projector, MPro150), Laser range sensor (URG-04LX by Hokuyo Electric Machinery), an USB camera (Logicool Inc., Qcam) and a pan-tilt unit (Directed Perception Inc., PTU-D46). The 3D mask and projector are mounted on the pan-tilt unit. The USB camera is wired on the top of the mask to detect frontal face of human and the laser range sensor is placed on the participant's shoulder level. To provide a communication channel between the hardware components of the system, there is a standard RS-232 serial port connection between the general purpose PC (Windows XP) and the pan-tilt unit. The LED projector projects CG generated eyes on the mask. Thus, the head can show nonverbal behaviors by its head movements and eye movements including blinking.

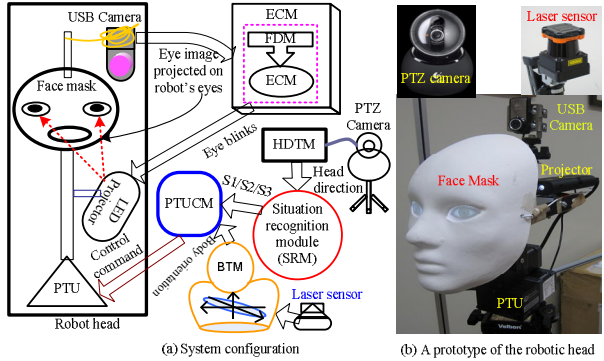

Figure 2. System consists of five modules: HDTM, SRM, BTM, ECM, and PTUCM (b) Prototype of the robotic head.

A PTZ camera (Logicool Inc., Qcam Orbit AF) is installed to track a human head and laser sensor track the human body. In the current implementation, PTZ camera and laser sensor are put on a tripod placed at an appropriate position to observe human body and head.

### 4.2 Software Configuration

The proposed system has five main software modules: the head detection and tracking module (HDTM), the body tracking module (BTM), the situation recognition module (SRM), the eye-contact module (ECM), and the pan-tilt unit control module (PTUCM). The last module controls the head movement and provides attention capture signals.

*Body Tracing Module (BTM):* A human body can be modeled as an ellipse [17]. We assume the coordinate system is represented with their X and Y axes aligned on the ground plane. Then, the human body model is consequently represented with center coordinates of ellipse [x,y] and rotation of ellipse (θ). These parameters are estimated in each frame by the particle filter framework [15]. We assume that the laser range sensor is placed on the participant's shoulder level so that the contour of his/her shoulder can be observed. When the distance data which captured by the laser range sensor is mapped on the 2D image plane, the contour of participant's shoulder is partially observed shown in Figure 3 (a).

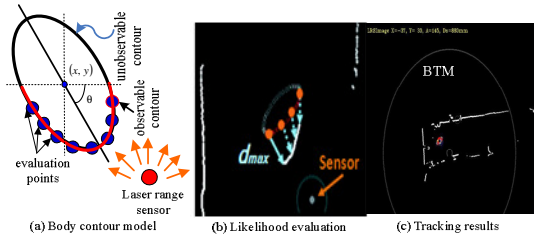

Figure 3. Results of BTM in terms of body position and orientation.

The likelihood of each sample is evaluated the maximum distance between evaluation points and the nearest distance data using the following equation.

$$\alpha = e^{-\left(\frac{d_{\max}^2}{\sigma_d}\right)}$$

where α is the likelihood score based on the laser image, $d_{max}$ is the maximum distance between evaluation points and the nearest distance data. At each time instance, once the distance image is generated from the laser image, each distance $d_n$ is easily obtained. $\sigma_d$ is the variance derived from $d_n$. Evaluation procedures are repeated for each sample. Conceptual images of evaluation process are shown in Figure 3 (b). We employ several points on the observable contour as the evaluation points to evaluate hypotheses in the particle filter framework. These points are changes depend on the relational position from the laser range sensor and the orientation of the model. Selection of evaluation points can be performed by calculating the inner product of normal vectors on the contour and its position vector from laser range sensor. A typical example of the result of the BTM is shown in Figure 3 (c). The BTM gives the body positions (x, y) of the human, distance between the human and laser sensor (D), and body orientation (θ). The results of the BTM (body orientation) send to the SRM to recognize OFOV situation and the robot adjust its head orientation based on the position of the human.

*Head Detection and Tracking Module (HDTM):* To detect, track and computes the direction of human head in real time (30 frame/sec), we use FaceAPI [9] by Seeing Machines Inc. It can measure 3D head position (x, y, z) and direction [yaw (α), pitch (β), and roll (γ)] within $3^0$ errors. One USB camera is placed in front of the human to track his/her face up to $\pm 90^0$. A snapshot of HDTM results has shown in Figure 4 (a). The results of the HDTM send to the SRM to classify the current viewing situations (CFOV, NPFOV and FPFOV) of the target person.

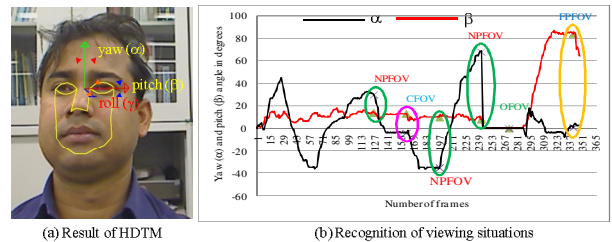

Figure 4. Recognition of situations based on head tracking information.

*Situation Recognition Module (SRM):* In order to recognize the existing situation (where the human is currently looking), we observe the head as well as body direction estimated by HDTM and BTM respectively. By extrapolating from the person's head/body information, the SRM determines which situation (CFOV, NPFOV, FPFOV, or OFOV) is exists between the robot and the human. By examining these situations, we found that the human head/body orientations are varying from one situation to other. Since the HDTM tracks within ±90$^0$ (right/left) only, therefore, while the human attend to OFOV situation, the system losses his/her head information, in that case, the robot recognize the current situation based on the body information (laser sensor can tracks up to 270$^0$). From the results of tracking modules, the system recognizes the following four situations of the target participant's in terms of yaw ($\alpha$), pitch ($\beta$) movements of head and/or body direction ($\theta$) respectively using a set of predefined rules. We have set the values for yaw, pitch and body directions by observing several experimental trials.

- **Central field of view (CFOV)**: recognized if the current head direction within $-10^0 \leq \alpha \leq +10^0$ and $-10^0 \leq \beta \leq +10^0$ and remains 30 frames in the same direction.
- **Near peripheral field of view (NPFOV)**: recognized if the current head direction within $-10^0 > \alpha \geq +70^0$ or $+10^0 \leq \alpha \leq 70^0$ and $-10^0 \leq \beta \leq +10^0$ and remains 30 frames in the same direction.
- **Far peripheral field of view (FPFOV)**: recognized if current head direction within $-0^0 > \alpha \geq +90^0$ or $+70^0 \leq \alpha \leq +90^0$ and $-10^0 \leq \beta \leq +10^0$ and remains 30 frames in the same direction.
- **Out of field of view (OFOV)**: recognized if the human looking to the opposite direction with respect to robot's direction. That means, the robot cannot capture the human face/head and current head direction within $\alpha = \beta = 0$ or body direction within $90^0 < \theta \leq +270^0$ or $-90^0 > \theta \geq -270^0$ and remains 30 frames in the same direction.

Fig. 4 (b) represents the results of SRM to recognize four situations.

*Eye Contact Module (ECM)*: The ECM mainly consists of two sub modules; FDM (Face detection module) and EBM (Eye blinking module). The robot continuously checks the target person's whether his/her face directed to the robot or not. In any situation, the robot considers that the human has responded against the robots' actions if s/he looks at the robot within expected times. In that case, the FDM uses the forehead camera to detect his/her frontal face [Figure 5(a)]. We use the face detector, which consists of cascaded classifiers based on AdaBoost and Haar-like features [2]. After face detection, the FDM sends the results to the EBM. The EBM producing eye blinks to let the person know that it is aware of his/her gaze. Since the eyes are CG images, the robot can easily blink the eyes in response to the human's gazing at it. Figs. 5(b)-(d) show some snapshots of a blinking action.

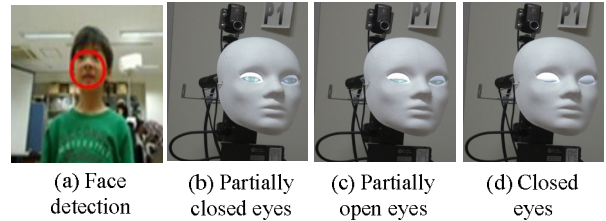

(a) Face detection   (b) Partially closed eyes   (c) Partially open eyes   (d) Closed eyes

Figure 5. Results of FDM and EBM.

*Pan-tilt Unit Control Module (PTUCM)*: In our proactive approach, the robot need to perform several actions (such as, head turning, head shaking, and uttering reference terms) to capture the human attention. All actions are performed by the pan-tilt unit with proper control signal coming from the several modules. Several properties of the robotic head are identified by experimental trials and summarized in Table 1.

Table 1. The properties of the robotic head.

| Items | Characteristics |
|---|---|
| Head turn (horizontal) | From -159$^0$ (left) to +159$^0$ (right) |
| Head turn (vertical) | From -47$^0$ (down) to +31$^0$ (top) |
| Eye turn | From -90$^0$ (left) to +90$^0$ (right) |
| Eye blinks rate | 1/seconds |
| Rotational speed | 300$^0$/second |
| Head tracking (error < 3$^0$) | From -90$^0$ (left) to +90$^0$ (right) |
| Body tracking (error < 6$^0$) | Up to 270$^0$ |
| Tracking distance | Up to 3 meters |
| No. of people tracking | 02 |

### 4.3 Behavioural Protocol of the Robot

This section describes the behavioural protocol of the robot. An eye contact event is executed by a finite-state-machine model as shown in Figure 6. In order to initiate the eye contact process, the robot begins to observe the current direction of the human's attention by tracking his/her head. After recognizing the viewing situation of the target human (TH), the robot usually turns its head first toward the TH, and commences shaking its head and then uttering reference terms (if necessary) to capture his/her attention. For the head turning (HT) action, we adjusted the pan speed of the pan-tilt unit at 120$^0$/second. For the head shaking (HS) action, the robot shook its head back and forth (±30$^0$) from its initial position. This meant that the robot turned its head 30$^0$ left and 30$^0$ right. The head-shaking speed was adjusted at 240$^0$/second. The system utter the terms ('excuse me') as the reference terms (RT). However, the robot waits about 4 seconds after each attempt for the TH to respond by looking in its direction[1].

If the robot is successful in attracting the TH's attention, the two agents will experience gaze crossing.

---

[1] Silences of more than 4 seconds become embarrassing because they imply a break in the thread of communication [21].

Thus, the robot considers the TH to have responded to its actions if he/she looks at the robot within the expected time frame. Otherwise, it considers the case as failure and initiates the interaction again. It is able to recognize whether this is so by detecting the front of his/her face in the camera image. After capturing the attention of TH, the robot performs a blinking action to display gaze awareness as an ensuring attention capture behavior.

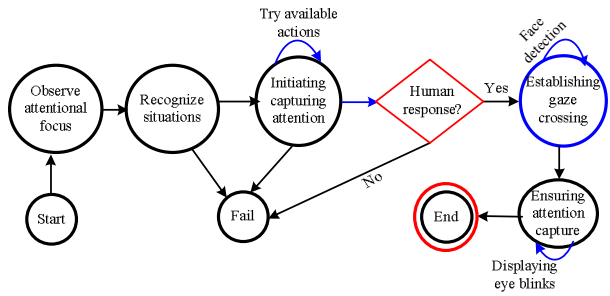

Figure 6. A pro-active behavioural model of eye contact.

## 5. An HRI Experiment

In order to evaluate the system, we performed an HRI experiment. In particular, the purpose of this experiment is to evaluate the effectiveness of our proposed robotic framework for making eye contact with the human while s/he was oriented toward a different viewing direction (i.e., when the robot and the human are not in face-to-face).

### 6.1 Participants

A total of 48 subjects (39 males, and 9 females) participated in the experiment. The average age of participants was 27.9 years (SD = 4.91). They were all graduate students at Saitama University, Japan. They were randomly assigned to one of the four conditions. There were 10 males and 2 females in the CFOV condition, 9 males and 3 females in the NPFOV condition, 11 males and 1 female in the FPFOV condition and 8 males and 4 females in the OFOV condition. Each participant experienced four types of behaviors of the robot one after another in four sessions in each viewing conditions (see in Section 6). Each session lasted approximately 120s. We deliberately concealed the primary purpose of our experiment. There was no remuneration for participants.

### 6.2 Experimental Design

As a low attention-absorption task we considered a scenario: 'watching paintings'. To prompt participants to look in various directions, we hung seven paintings (P1-P7) on the wall at the same height (just above the eye level of the participants). These paintings were placed in such a way that, when observed from a participant's sitting position, they covered their whole field of view (close to $180^0$). To produce the stimuli, we prepared two robotic heads with the same appearance. The mere existence of such robots in an environment may prompt participants to be attracted to them because of their human-face-like appearance, even if they do not perform any actions [6]. One was a static robot (SR), which was stationary at all times. The other was a moving robot (MR). Initially MR is static and is looking in a direction not toward the human face. Two robots were placed in the participant's left and right monocular fields of view. Participants' head direction would change while watching these paintings. The roles of the left and right robotic heads were exchanged randomly so that the number of participants experienced each case could be almost the same. Two video cameras were placed in appropriate positions to capture all interactions. Figure 7 shows the schematic setting of the experiment.

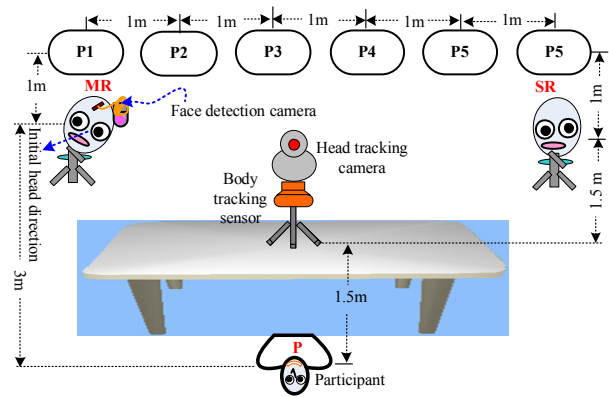

Figure 7. Schematic settings of the experiment.

### 6.3 Experimental Procedure

Our intention was to let the participants evaluate the various behaviors of the robot as it attempted to acquire their attention when they were not initially looking in its direction. For this purpose, a single participant was asked to sit down on chair and asked to look around at the paintings. We let the participant to watch the paintings. The robot tracks the participant and hence the MR did not perform any action during first 60 s of the interaction.

During observation of paintings, MR shows all actions (during last 60s) to the participant one after another in each viewing condition to capture his/her attention. If the participant looks at MR within expected time frame (i.e., 4s), the robot considers that s/he has been attracted. In this experimental scenario, if the target participants did not gaze at the robot within the expected time frame following the robot's actions, then the robot considered the case to be a failure. We videotaped all sessions to analyze human behaviors. Figure 8 shows an experimental scene while interacting with the robots.

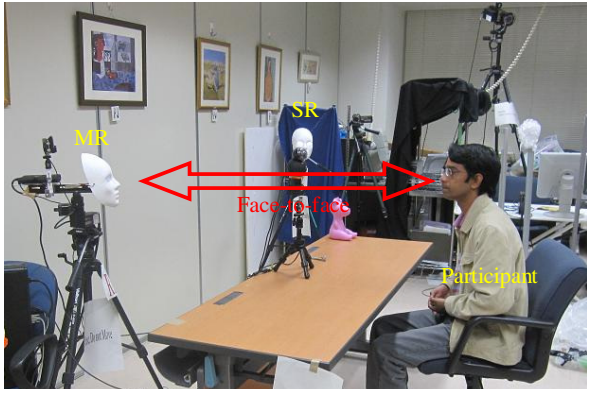

Figure 8. An experimental scene in which the participant attracted and looked at the moving robot (MR).

## 6.4 Experimental Conditions

The robot tried to gain the participant's attention while s/he was looking at different paintings so that it could obtain data for four types of viewing situation. The robot shows all actions each viewing situation. Thus, we adopted four viewing situation and four behavioral action conditions. They were defined as follows:

  (a) **Viewing situations:** By our observation, we seen that the robot recognition the situation as **CFOV** when the participant looking at the painting P1, as **NPFOV** when the participant looking at the picture P2/P3, as **FPFOV** when s/he looking at the picture P4/P5. However, it recognized as **OFOV** while looking at the picture P6.
  (b) **Actions:** In all actions, the robot turns its head in another direction after waiting 3s.

- Method 1 (**M1**): The robot always applies HT action to attract the participant's attention whatever the situation is. If the target person is looking at the robot, it blinks its eyes about 3s.
- Method 2 (**M2**): The robot turns its HT to gain the participant's attention toward it and then shaking head (if necessary) in all situations. If the target person is looking at the robot, it blinks its eyes about 3s.
- Method 3 (**M3**): In order to capture the participant's attention, the robot applies HT first, and then HS. The robot applies RT action only if previous actions are failed to capture the participant's attention. If the target person is looking at the robot, it waits about 3s but does not display any eye blinking action.
- Method 4 (**M4**): This is our proposed robot. The details description of the robot has described in Section 4.3. The robot blinks its eyes about 3s if it gained the participant's attention toward it.

## 6.5 Measures

The measures of this experiment will perform in quantitative and subjective ways.

### 6.5.1 Quantitative Measures

By observing the experimental videos, we measure the following items:
- *Success ratio*: refers to the ratio between the number of cases where participants looked at the robot in response to its action ($N_L$) and the total number of cases ($N_A$).
- *Gazing time*: We measure the total time spent by gazing at the robot in each method by observing the experimental videos. This time is measured from the beginning of gaze crossing action of the robot to the end of the participant's looking at it before turning head to another direction.

### 6.5.2 Subjective Measures

We asked participants to fill out a questionnaire after interactions with the robots were complete. The measurement was a simple rating on a Likert scale of 1 to 7, where 1 stands for the lowest and 7 for the highest. The questionnaire had the following items:
- **Attention attraction**: Did you feel that behaviors of the robot captured your attention?
- **The feeling of being making eye contact**: Did behaviors of the robot created your feeling of making eye contact?
- **Overall evaluation**: How effective the robot for making eye contact?

## 6.6 Results

The experiment conducted was a 4×4 mixed-model design. For within-participant factor (*action*), all participants interacted with four actions of the robot (M1, M2, M3, and M4) and for between-participant factor (*viewing situation*) one group of participant were experienced the four actions in one of the four viewing situations (CFOV, NPFOV, FPFOV, and OFOV). We observed a total of 192 (12 [*participants*] × 4 [*actions*] × 4 [*situations*]) interactions.

### 6.6.1 Quantitative Measures

Table 2 summarizes mean and standard deviation (SD) of participant's response with respect to the robot's behaviours in each viewing situation.

Table 2. Summary of success ratio of capturing participant's attention against each action in different viewing conditions.

| Actions | Viewing Situations | | | |
|---|---|---|---|---|
| | CFOV | NPFOV | FPFOV | OFOV |
| | Mean (S. D) | Mean (S. D) | Mean (S. D) | Mean (S. D) |
| **M1** | 0.92 (0.29) | 0.84 (0.39) | 0.08 (0.29) | 0.08 (0.29) |
| **M2** | 1.0 (0.0) | 0.92 (0.29) | 0.84 (0.39) | 0.16 (0.39) |
| **M3** | 1.0 (0.0) | 0.92 (0.29) | 0.92 (0.29) | 0.92 (0.39) |
| **M4** | 1.0 (0.0) | 0.92 (0.29) | 0.92 (0.29) | 0.92 (0.29) |

A two-way repeated-measure of ANOVA was conducted for the success ratios. A significant main effect were revealed in the action factor ($F(3,176)=25.6$, $p<0.001, \eta^2=0.18$) and viewing situation factor ($F(3,176)=26.28$, $p<0.001$, $\eta^2=0.19$). The interaction effect between the movement and viewing situation was significant ($F(9,176)=8.3$, $p<0.01$, $\eta^2=0.18$). Figure 9 also illustrates these results.

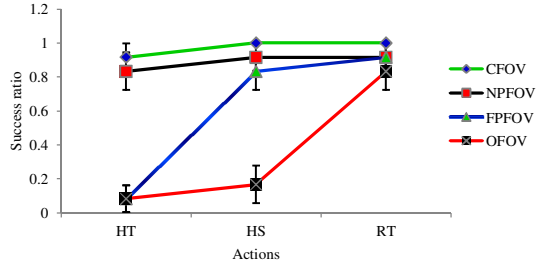

Figure 9. Mean values of success ratio of the robot in different actions. Error bars indicates the standard deviation.

The significant interaction effect between viewing situation and action suggests that the success ratios for different methods are affected by the viewing situation factor. Post hoc tests for the viewing condition revealed significant differences between pairs (CFOV and FPFOV: $p<0.01$, CFOV and OFOV: $p<0.01$, NPFOV and FPFOV: $p<0.01$, NPFOV and OFOV: $p<0.01$) but there was no significant difference between CFOV and NPFOV for M1 action. That means M1 is effective for CFOV and NPFOV situations. Moreover, multiple comparisons with the Bonferroni method were conducted among four action parameters for each viewing situation condition. For CFOV and NPFOV conditions, no significant differences were found between any action pairs (i.e., M1 and M2, M2 and M3, M3 and M1, M3 and M4). In these conditions, all actions are equally effective to capture participant's attention toward the robot. In particular, success ratios are higher for HT action both in CFOV and NPFOV situations than in FPFOV and OFOV situations. That means, HT action is sufficient to capture the human attention while s/he was perceived the robot in his/her CFOV or NPFOV situation.

Concerning M2, Post hoc tests for the viewing condition revealed significant differences between pairs (CFOV vs. OFOV: $p<0.01$, NPFOV vs. OFOV: $p<0.01$, FPFOV vs. OFOV: $p<0.01$) but no significant differences were found for the other pairs (CFOV vs. NPFOV, CFOV vs. FPFOV, NPFOV vs. FPFOV). That means M2 is effective for CFOV, NPFOV and FFOV situations but not effective in OFOV situation. Multiple comparisons with the Bonferroni method for FPFOV condition shows significant differences between the actions pairs (M1 vs. M2: $p<0.0001$, M3 vs. M1, and M4 vs. M1: $p<0.0001$). No significant differences were found between pairs (M2 and M3, M3 and M4, M2 and M4). In particular, HS action of the robot achieved the higher success ratio than HT in FPFOV condition and most of the participants responded to the robot after HS action. Thus, the robot should use more strong actions in FPFOV viewing condition to gain the participants' attention.

Concerning M3 and M4, Post hoc tests for the viewing condition revealed no significant differences between all pairs which mean that M3 and M4 are effective for all situations in capturing participant's attention due to their same action plans. For OFOV condition, significant differences were found between the actions pairs (M2 vs. M3: $p=0.0002$, and M3 vs. M1: $p<0.0001$). No significant difference was between pairs (M1 and M2, M3 and M4). This means that the RT action of the robot achieved the higher success ratio than HT and HS in the OFOV condition. Thus, it cannot be possible to capture the human attention by any kind of physical action when the robot is exist in such a position from where s/he cannot see the robot. In that case, using voice or sound action should be used to capture people attention.

For overall evaluation, we conducted multiple comparisons with the Bonferroni method that showed significant differences between M1 and M4 ($p = 0.001$), between M2 and M4 ($p =0.01$), between M4 and M1 ($p < 0.001$). Results also revealed that a substantial 93% of target participants' attention was captured by the proposed method, while only 48% and 73% of their attention was captured by methods 1, and 2 respectively. Figure 10 also shows these results. Results mean that the capturing attention performance of the robot is clearly more effective compared to the other two methods, in terms of producing a higher success ratio, when it employs the HT, HS, and RT actions.

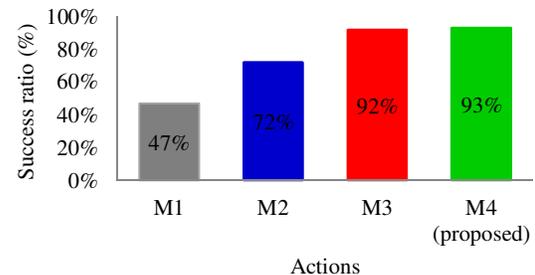

Figure 10. Overall success ratio of capturing attention.

We calculate the total time that the participants were spent to look at the robot in M3 and M4 after meeting face-to-face (Table 3). We compare only M3 and M4 due to their similar attention capturing action plan. Results indicate that the participant looks significantly longer in proposed method (2.46 seconds) than the other method (1.13 seconds). ANOVA analysis showed that there are significant differences that the participants spending to gaze at the robot in each method ($F(1,95)=445.9$, $p<0.0001$, $\eta^2=0.8$).

Table 3 Results of total time spent on gazing at two robots.

|  | Time (seconds) | | F (p) |
| --- | --- | --- | --- |
|  | M4 | M3 |  |
| Mean | 2.51 | 1.1 | 445.9 (<0.0001) |
| Var | 0.13 | 0.01 |  |

### 6.6.2 Subjective Measures

Table 4 shows the participants response on each question.

Table 4: Results of subjective measure in terms of mean (M) and standard deviation (SD). S.L means the significant level with probability (p).

|  | Actions | Attention capture | The feeling of being making eye contact | Overall evaluation |
| --- | --- | --- | --- | --- |
| **M (SD)** | M1 | 2.6 (1.63) | 2.75 (1.76) | 1.34 (0.48) |
|  | M2 | 3.45 (1.35) | 4.5 (0.84) | 1.79 (0.65) |
|  | M3 | 5.4 (0.61) | 3.02 (0.84) | 2.3 (0.75) |
|  | M4 | 5.5 (0.55) | 5.29 (0.58) | 5.35 (0.60) |
| **S. L (p)** | M1 vs. M4 | 2.6 (<0.0001) | 2.75 (<0.0001) | 1.34 (<0.0001) |
|  | M2 vs. M4 | 2.6 (<0.0001) | 2.75 (=0.0002) | 1.34 (<0.0001) |
|  | M3 vs. M4 | 2.6 (=0.59) | 2.75 (<0.0001) | 1.34 (<0.0001) |

Concerning in capturing participants' attention, ANOVA analysis shows that there are significance differences among action condition ($F(3,191)=79.08$, $p<0.0001$, $\eta^2 = 0.5$) [Figure 11]. Multiple comparison with Bonferroni method shows a significant differences between M1 and M4 ($p<0.0001$) and between M2 and M4 ($p<0.001$) but shows no significant difference between the robot with blinks and without blinks conditions (i.e. M3 and M4: $p=0.59$). This happens due to the same attention capturing behaviours of robot in two conditions.

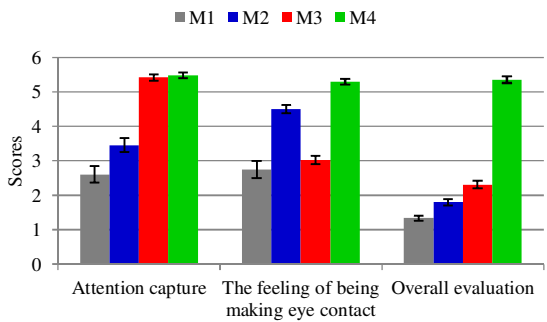

Figure 11. Questionnaires responses of participants. Error bars indicates the standard deviation and *** means significant differences.

In the case of feeling of being making eye contact, ANOVA analysis shows a significant difference among action conditions ($F(3, 191) = 45.04$, $p < 0.001$, $\eta^2 = 0.44$) [Fig. 11]. Multiple comparison with the Bonferroni method showed significant differences between methods 1 and 4 ($p < 0.0001$), between methods 2 and 4 ($p = 0.0002$), between methods 3 and 4 ($p < 0.001$) respectively. This result revealed that the participant's impressions are greatly affected by the eye blinking behaviors of the robot and this behavior produced a better feeling of making eye contact.

Concerning the overall evaluation, a significant main effect was found ($F(3, 191) = 357.4$, $p < 0.0001$, $\eta^2 = 0.86$) using ANOVA analysis. Fig. 11 also illustrates this result. Participant rated more the robot with blinks condition (M4) [Mean score =5.35] than the robot with no blinks condition (M3) [Mean score=2.3]. Multiple comparison with the Bonferroni method also showed significant differences between M1 and M4 ($p < 0.0001$), between M2 and M4 ($p < 0.0001$), between M3 and M4 ($p < 0.0001$) respectively. Thus, the results reveal that the proposed system is more preferable than the other methods to make eye contact with the participants.

## 6. Discussion

In proactive approach, the robot should capture the target human attention first for establishing eye contact. Our purpose is to develop a robot that can make eye contact with a particular human while avoiding attracting other people's attention as much as possible. Thus, the robot should consider the current situation of intended people with whom it would like to start communicating and try to apply an appropriate action to that situation. For this purpose, we propose an eye contact process consisting of capturing attention and ensuring attention capture. To initialize an eye contact episode, the robot should start with a weak action to avoid attracting other people than the target person and use stronger actions when the situation becomes tougher. This is the basic design concept of our robot. From the survey of psychology and HRI literatures, we chose turning the head (to look at the person) as the weakest action. We determined to use head shaking if the robot cannot attract the target person's attention and use reference terms if the robot captured the target person in its out of field of view condition. We have confirmed through experiments that our design concept can be useful to realize such robots that can captured a particular person as selectively as possible.

Blinking actions strengthens the feeling of being looked at and it can be used to convey an impression more effectively and colorfully understanding of human social behavior. Experimental results have also confirmed eye blinking actions proved helpful to relay to the target that the robot was aware of his/her gaze.

Making eye contact pro-actively is an important social phenomena and prerequisite in several social functions such as engagement, initiating conversation, shared attention, and so on. The robot may use proactive approach for making eye contact in several contexts (i.e., information providing services, providing route direction, salesperson, tutoring services, and so on) that are demanded such kinds of social functions.

## 7. Conclusion

The primary focus of our work is to develop a robot that can make eye contact with a particular person's by nonverbal means. For this purpose, we have proposed a proactive approach of eye contact that consists of two phases including capturing attention, and ensuring attention capture. Although there may be various non-verbal behaviors, we incorporated head movements, reference terms, and eye blinking in respective phases. We have shown that our method can functioning to establish a eye contact event with the target human in a situation where s/he is not initially looking toward the robot (in particular, we have considered three such situations namely, NPFOV, FPFOV, and OFOV) and is involved in a task that does not demand much attention. If the participant is paying attention to a particular object or talking with another person, the robot needs to use some other actions. There are other behaviors to capture attention, such as eye movement, waving or the combination of verbal and nonverbal actions. These are left for future work.